\pdfoutput=1
\documentclass[twocolumn,english,aps,superscriptaddress,prb]{revtex4-1}

\usepackage{babel}
\usepackage{amsmath}
\usepackage{amssymb}
\usepackage{graphicx}

\begin{document}

\title{Floating phase versus chiral transition in a 1D hard-boson model
}

\author{Natalia Chepiga}
\affiliation{Department of Physics and Astronomy, University of California, Irvine, CA 92697, USA}
\author{Fr\'ed\'eric Mila}
\affiliation{Institute of Physics, Ecole Polytechnique F\'ed\'erale de Lausanne (EPFL), CH-1015 Lausanne, Switzerland}

\date{\today}
\begin{abstract} 
We investigate the nature of the phase transition between the period-three charge-density wave and the disordered phase of a hard-boson model proposed in the context of cold-atom experiments. Building on a density-matrix renormalization group algorithm that takes full advantage of the hard-boson constraints, we study systems with up to 9'000 sites and calculate the correlation length and the wave-vector of the incommensurate short-range correlations with unprecedented accuracy. We provide strong numerical evidence that there is an intermediate floating phase far enough from the integrable Potts point, while in its vicinity, our numerical data are consistent with a unique transition in the Huse-Fisher chiral universality class. 
\end{abstract}

\maketitle


The identification of the universality class of phase transitions is one of the most important aspects of both classical and quantum physics. In the presence of a broken symmetry, simple symmetry arguments often allow one to guess the universality class of a transition (Ising, 3-state Potts, etc.) depending on the degeneracy of the broken symmetry state. There are however cases where this is not sufficient. A prominent example is the commensurate-incommensurate transition in the case of a commensurate phase with three types of domains. As first proposed by Huse and Fisher\cite{HuseFisher}, if domain walls between different phases have different properties, this introduces a chiral perturbation (the sequence say $A | B | C$ is not equivalent to its mirror image $A | C | B$, where $A$, $B$, $C$ refer to different domains), and if this perturbation is relevant, the transition can only be in the 3-state Potts universality class at an isolated point where the perturbation vanishes. Away from that point, there are three possibilities: (i) There is still a unique transition, but it belongs to a new universality class called {\it chiral}; (ii) There is a critical incommensurate intermediate phase called a {\it floating} phase; (iii) The transition is first order. The investigation of this problem has been a hotly debated issue in the eighties in the context of solid-on-solid models of adsorbed layers\cite{ostlund,huse,HuseFisher, Selke1982,CENTEN1982585,HOWES1983169,haldane_bak,schulz,Duxbury,HuseFisher1984,bartelt}, and the chiral Potts model has been further studied since then\cite{Baxter1989,albertini,mccoy,cardy,fendley_parafermions,hughes,sachdev_dual}. Experimental evidence of the chiral melting of Ge(113) and Si(113) $3 \times 1$ phases has been reported in the early nineties\cite{SelkeExperiment}.

\begin{figure}[t!]
\includegraphics[width=0.45\textwidth]{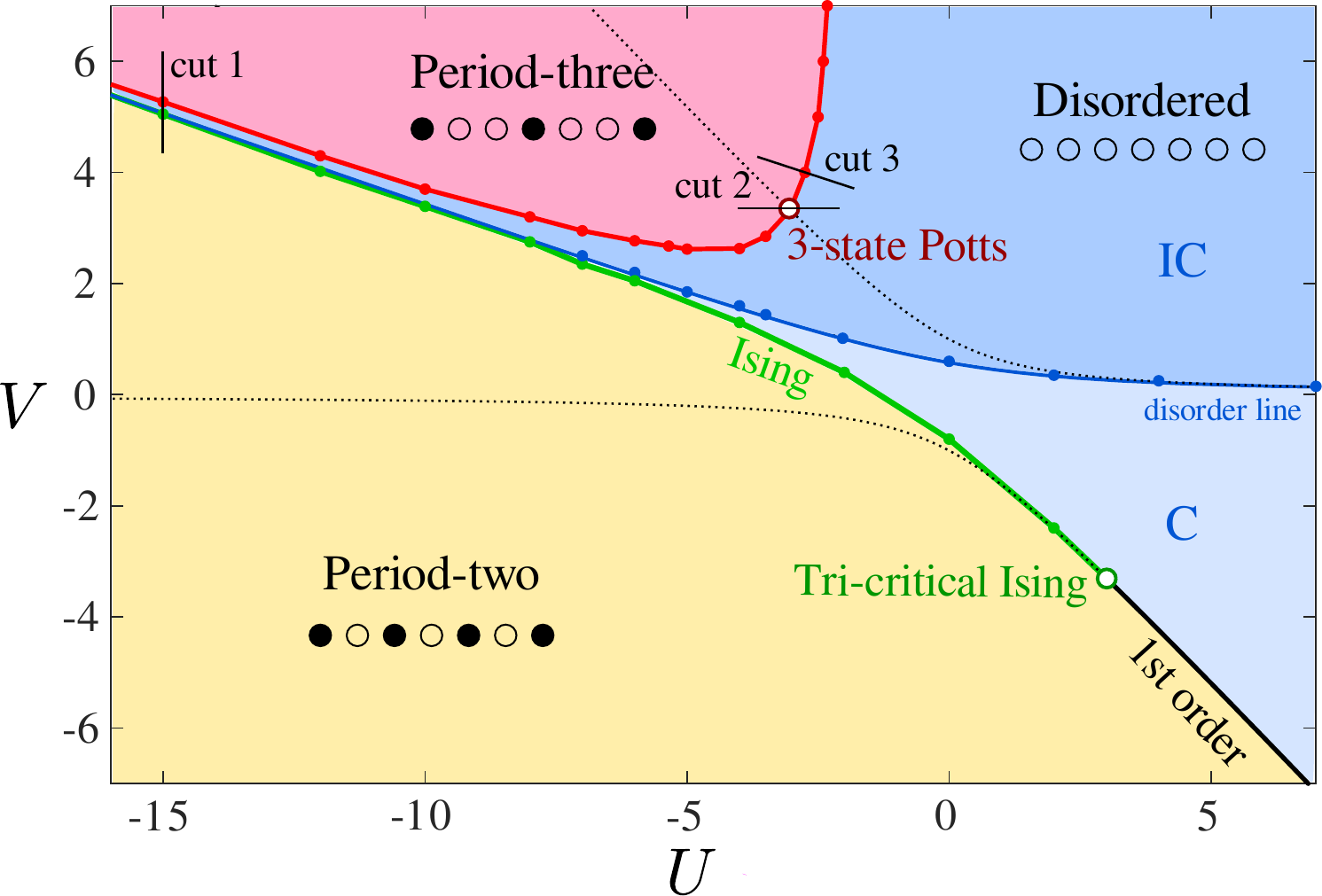}
\caption{Phase diagram of the hard-boson model of Eq.\ref{eq:hard_boson}, with three main phases: two ordered ones with period two and three, and a disordered one with incommensurate short-range correlations (IC) above the disorder (blue) line and commensurate (C) ones below. The transition out of the period-two phase is first order (black line) below a tri-critical Ising point (open green circle) and in the Ising universality class beyond it (green line). The transition out of the period three phase is expected to be in the 3-state Potts universality along the integrable line (dotted line), in the chiral Huse-Fisher universality class close to it, and through an intermediate critical phase with incommensurate correlations far from it (see main text). The width of this phase is smaller than that of the red line.
Thin black lines indicate the three cuts used in Fig.\ref{fig:correlation_length}(a-i). 
}
\label{fig:phase_diagram}
\end{figure}

The issue has been recently reopened by Fendley et al\cite{fendley} in the context of a 1D quantum model of trapped alkali atoms\cite{sachdev_girvin} also relevant for recent experiments on Rydberg states\cite{Bernien2017,Lesanovsky2012} described by the Hamiltonian:
\begin{equation}
  H=\sum_j \left[ -w(d_j^\dagger+d_j)+Un_j+Vn_{j-1}n_{j+1}\right],
  \label{eq:hard_boson}
\end{equation}
In this model, $d_j^\dagger$ and $d_j$ are creation and annihilation operators of {\it hard bosons} defined by the constraints $n_j(1-n_j)=0$ (no double occupancy, as for hard-core bosons) and $n_jn_{j+1}=0$ (bosons cannot sit on neighboring sites). As shown by Fendley et al\cite{fendley} and confirmed by our systematic investigation of the whole parameter space with DMRG simulations, the phase diagram of this model (see Fig.\ref{fig:phase_diagram}) consists of three main phases : two ordered phases (period two and period three respectively), and a disordered phase. There is a disorder line in the disordered phase accross which short-range correlations become incommensurate. 
Correlations are commensurate close to the period-two phase, and the transition out of this phase turns from Ising to first-order through a tricritical Ising point. Close to the period-three phase, the correlations are incommensurate, and the melting of the period-three phase is an example of a commensurate-incommensurate transition with a relevant chiral perturbation. The model has an integrable line (dotted line) along which the transition is in the 3-state Potts universality class, and the main open issue is the nature of the melting away from it. In the limit  $U\rightarrow - \infty$, Bethe ansatz results have shown that there has to be an intermediate floating phase, and the absence of indication of an additional transition has led the authors of Ref.\onlinecite{fendley} to suggest that, coming from that side, an intermediate phase might be present up to the Potts point. More recently, this conclusion has been challenged by Samajdar et al\cite{samajdar}, who see no reason to discard the original scenario put forward by Huse and Fisher \cite{HuseFisher} with a Potts point flanked by chiral transitions, and who provided numerical evidence of a dynamical exponent larger than 1 on the other side of the Potts point, in agreement with a chiral phase. Note that on the scale of Fig.\ref{fig:phase_diagram}, the intermediate phase (if any) is narrower than the line width.

To investigate the competition between a chiral transition and a floating phase, the most direct evidence relies on the behavior of the wave-vector and the correlation length close to the transition. For the 3-state Potts universality class, the wave-vector $q$ is expected to approach $2 \pi/3$ with an exponent  $\bar \beta=5/3$, as first shown by Baxter and Pearce \cite{Baxter_Pearce,HuseFisher1984}, while the correlation length $\xi$ is expected to diverge with an exponent $\nu=5/6$. In the ordered phase, the correlation length is also expected to diverge with an exponent $\nu'=5/6$. By contrast, if the transition is chiral, it has been predicted by Huse and Fisher that $\bar \beta=\nu$, so that $|q - 2\pi/3| \times \xi$ tends to a constant at the transition. Besides, there is still a unique transition, and $\nu'=\nu$. The presence of an intermediate floating phase can also be clearly identified: Coming from the disordered phase, the correlation length diverges at a Kosterlitz-Thouless (KT) transition\cite{Kosterlitz_Thouless} at a point where $q$ is still incommensurate, and $q$ reaches $2\pi/3$ at a subsequent Pokrosky-Talapov (PT) transition\cite{Pokrovsky_Talapov}, with an exponent $\bar \beta=1/2$. In addition, the correlation lengths behave very differently: at the KT transition, the correlation length diverges as $\xi\propto \exp(\text{C}/\sqrt{g_{KT}-g})$, where $C$ is a constant and $g$ a coordinate along the path in parameter space, while at the PT transition coming from the ordered phase, it diverges with an exponent $\nu'=1/2$. As we shall see, such an analysis requires to have access to system sizes that are beyond the scope of standard density-matrix renormalization group (DMRG) algorithms\cite{dmrg1,white1993,dmrg2,dmrg3,dmrg4} for hard-core bosons, which can typically handle hundreds but not thousands of sites. 

In this Letter, building on the exact mapping of the model of Eq. \ref{eq:hard_boson} onto a quantum dimer model on a ladder\cite{chepiga}, we develop a DMRG algorithm that takes the hard-boson constraints explicitly into account and thus takes full advantage of the fact that the Hilbert space only grows as Fibonacci number\cite{sierra}. This allows us to reach very large system sizes (routinely 4'800 sites, up to 9'000 sites occasionally) and to access the scaling properties of the wave-vector and of the correlation length, hence to investigate the competition between a direct chiral transition and an intermediate critical floating phase. All simulations have been performed with open boundary conditions, and the wave-vector and the correlation lengths have been obtained by fitting the density-density correlations with Ornstein-Zernike\cite{ornstein_zernike} (see Supplemental Material\cite{SM} for details).

\begin{figure*}[t!]
\includegraphics[width=\textwidth]{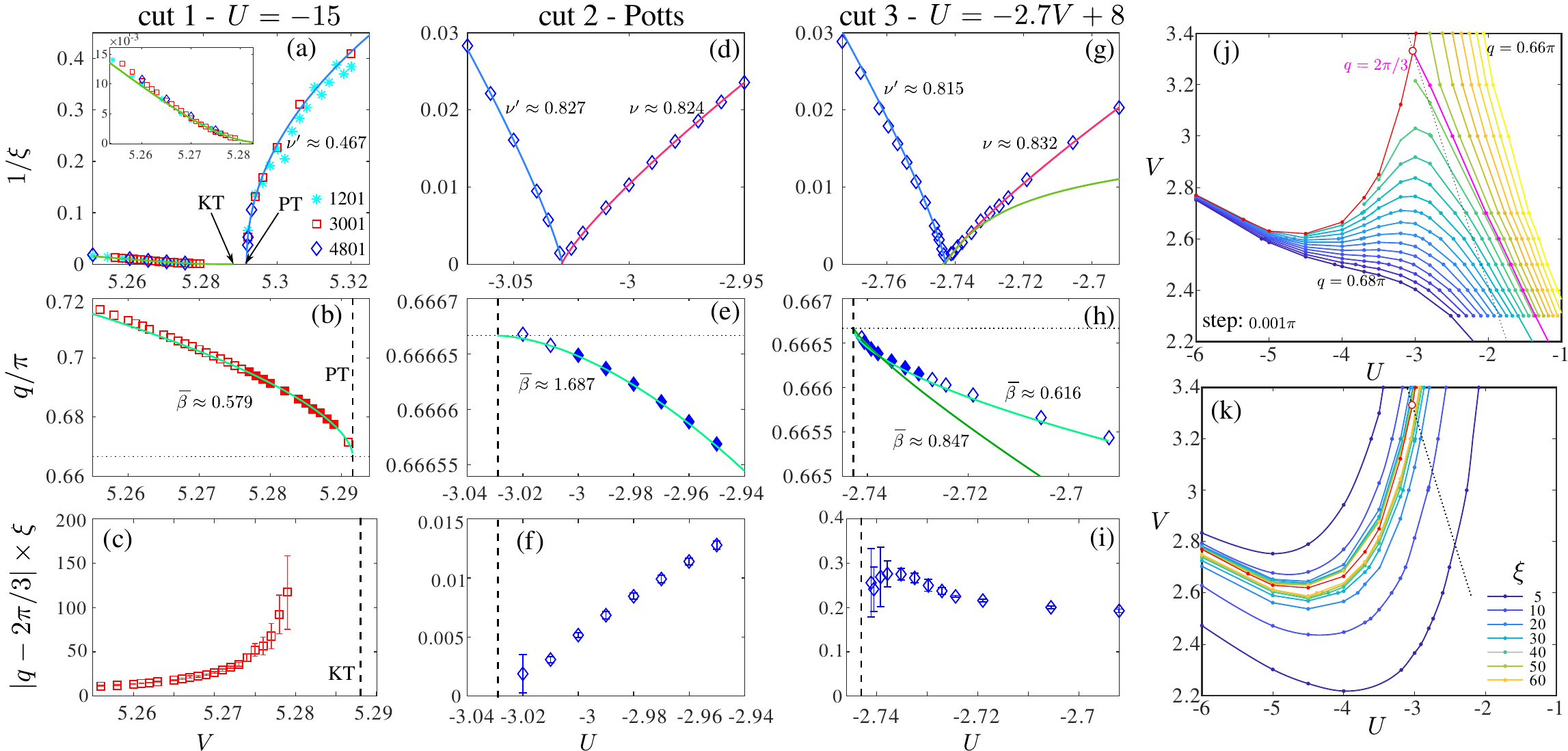}
\caption{(a-i) Inverse correlation length $1/\xi$, wave vector $q$ of the incommensurate correlations, and product  $|q - 2\pi/3| \times \xi$ for three cuts shown in Fig.\ref{fig:phase_diagram}.
(a-c) Vertical cut at $U=-15$; (d-f) Horizontal cut through the Potts point $V\simeq 3.3302$; (g-h) Cut along the $U=-2.7V+8$ line just above the Potts critical point.
Inside the period-three phase, the correlation length is fitted with a power-law with critical exponent $\nu^\prime$ (dark blue lines). In the disordered phase, the correlation length is fitted either with a power-law with critical exponent $\nu$ (pink line), or with the KT form $\xi\propto \exp(\text{C}/\sqrt{g_{KT}-g})$, where $g$ is the coordinate along the cut (green lines). 
The approach of the wave-vector $q$ to $2\pi/3$ is fitted with a power law with critical exponent $\bar \beta$, assuming the critical coupling determined from the correlation length in the ordered phase (filled symbols show the points used for the fit, except for the green line with exponent $\bar \beta \simeq 0.847$ in panel (h), for which only the points closest to the transition have been used).
(j-k) Constant-$q$ and $\xi$ plots in the vicinity of the boundary of the period-three phase.}
\label{fig:correlation_length}
\end{figure*}

The gross features of the divergence of the correlation length and of the approach of the wave-vector to $2\pi/3$ are illustrated in Figs. (2j-k). While the correlation length increases smoothly along the border, the wave-vector is larger than $2\pi/3$ below a separatrix defined by $q=2\pi/3$ and smaller than $2\pi/3$ above it, leading to turning points close to the Potts point. Finite-size effects can thus be expected to be larger for the wave-vector than for the correlation length, especially below the Potts point, where the turning point is rather acute. 

Let us now look in more details at the vicinity of the transition line. We start by discussing three specific points where three rather different behaviors are observed.

The Potts nature of the critical point along the integrable line at $U=\varphi^{-5/2}-\varphi^{5/2}\simeq -3.0299, V=\varphi^{5/2}\simeq 3.3302$, where $\varphi=(\sqrt{5}+1)/2$ is the golden ratio, is well established from the exact solution\cite{fendley}. To benchmark our method, we have studied a horizontal cut that goes through this Potts point. The results shown in Fig. 2(d-f) are fully consistent with the theoretical predictions $\bar\beta=5/3$ and $\nu=\nu'=5/6$. 

Next, we consider a vertical cut far from the Potts point at $U=-15$. The numerical results are summarized in Figs. 2a-c. The most striking difference with the previous cases is the behavior of the correlation length. In the disordered phase, the inverse correlation length vanishes with a vanishing slope, whereas in the ordered phase, it vanishes with an exponent $\nu'\simeq 0.5$. This asymmetry strongly suggests that we are not facing a unique transition any more, but two transitions of different nature. Fits of the correlation length assuming that the transition is KT in the disordered phase
and PT in the ordered one leads to $V_{KT}\simeq 5.288 <V_{PT}\simeq 5.291$, confirming the presence of two transitions with a narrow intermediate phase. The width of this phase is very small, but interestingly enough, it is compatible with the Bethe ansatz prediction in the $U\rightarrow -\infty$ limit\cite{comment1}. The behaviour of the wave-vector further confirms this conclusion: $q$ reaches $2\pi/3$ at the PT transition and with an exponent $\bar \beta\simeq 0.579$, in reasonable agreement with the PT prediction $\bar \beta= 1/2$. 
Finally, we have also calculated the central charge between $V_{KT}$ and $V_{PT}$, and it agrees with 3\% with the Luttinger liquid prediction $c=1$ for the intermediate floating phase\cite{SM}. 

Let us now concentrate on a cut slightly above the Potts point defined by the equation $U=-2.7 V+8$. The numerical results are summarized in Fig. 2(g-i). The correlation lengths measured in the disordered and ordered phases behave essentially as when crossing the Potts point. They are consistent with a single transition point at which both diverge, and the exponents $\nu\simeq 0.832$ and $\nu'\simeq 0.815$ are equal to a good accuracy. By contrast to the Potts case however, the wave-vector approaches $2\pi/3$ with an exponent $\bar \beta<1$ (see the change of concavity between Figs. 3e and 3h). So the transition is clearly not in the Potts universality class. At the same time, the behavior of the correlation length in the disordered phase is inconsistent with a KT transition: There is no sign of a change of curvature in $1/\xi$, and if we nevertheless try to fit it with the KT scaling, we get a transition point inside the ordered phase. The presence of an intermediate phase is thus rather unlikely.

Taken together, these various pieces of information point to the Huse-Fisher chiral universality class as the only possibility. This is further confirmed by our numerical results for $|q - 2\pi/3| \times \xi$ shown in Fig. 2(i). Within the error bars, they are consistent with a finite, non-zero limit at the transition, as predicted by Huse and Fisher\cite{HuseFisher}. This should be contrasted with the vanishing limit at the Potts point (see Fig. 2(f)), and with the divergence at the KT transition when there is an intermediate phase (see Fig. 2(c)). Note that a direct determination of the exponent $\bar \beta$ is tricky.  If all points between $V_c$ and say $V_c-0.01$ are included, a fit of $q$ with a power law yields $\bar \beta \simeq 0.62$. If however only the points very close to the transition are included, the exponent takes the value $\bar \beta \simeq 0.847$, in reasonable agreement with the prediction $\bar \beta = \nu$. This crossover is probably due to the proximity of a PT transition.

\begin{figure}[t!]
\includegraphics[width=0.48\textwidth]{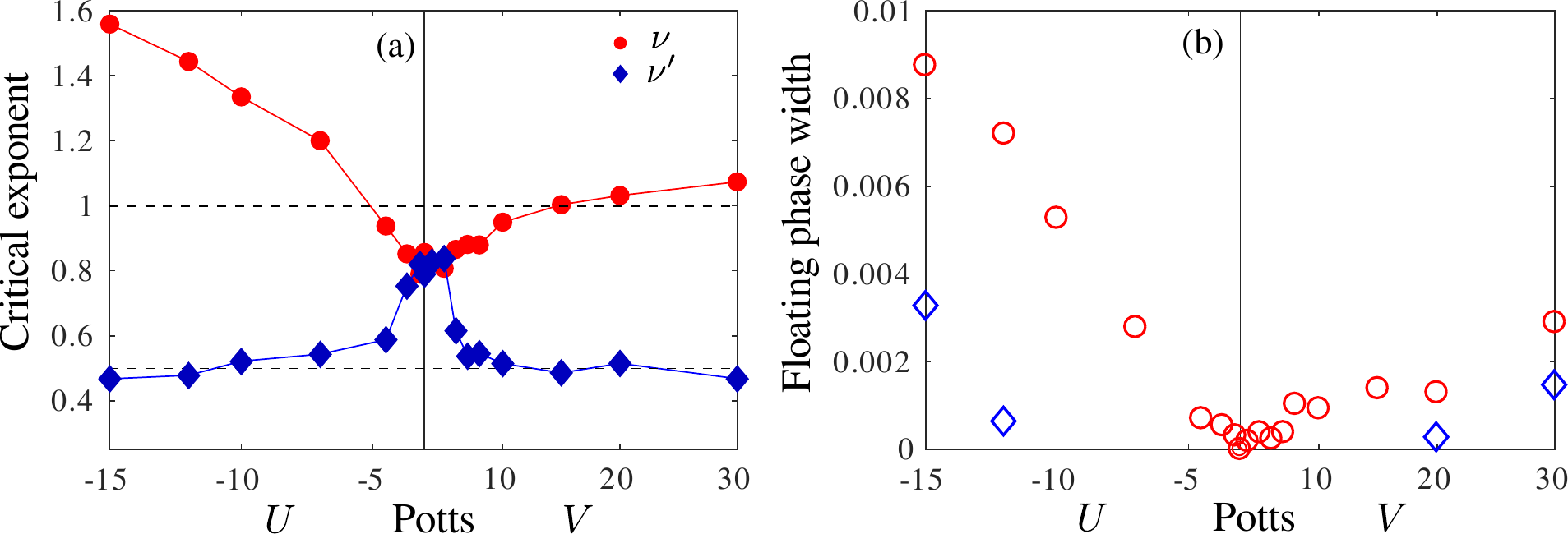}
\caption{(a) Critical exponents obtained by fitting the correlation length with a power law in the disordered phase ($\nu$, red circles) and in the ordered phase ($\nu'$, blue diamonds); (b) Splitting between the critical couplings deduced from fitting both correlation lengths with power laws (red circles), or from fitting that of the ordered phase with a power law and that of the disordered phase with with the KT prediction (blue diamonds). 
}
\label{fig:exponent}
\end{figure}

Now, let us look at the trends along the critical line. The main results are presented in Fig. 3. Details are provided in the Supplemental Material\cite{SM}. In Fig. 3(a), we show the value of the exponents $\nu$ and $\nu'$ along the transition line obtained by candid fits of the correlation length in the disordered phase resp. in the ordered phase with power laws. Of course, the fit with a power law is meaningless if the transition is KT, but this graph is very useful to demonstrate that there are two very different parameter ranges: close to the Potts point (say $U>-4.5$ or $V<6$), the two exponents are very close to each other, 
while outside this range, they take very different values: $\nu'\simeq0.5$, consistent with a PT transition, and $\nu>1$, inconsistent with a chiral transition\cite{cardy}. In Fig. 3(b), we show the difference between the critical values of the coupling constants deduced from power law fits of both transitions (red symbols), and from fits with KT on one side and a power law on the other side (blue symbols), whenever a KT fit was conclusive. The difference between the critical values deduced from power-law fits is below our precision close to the Potts point as long as $\nu \simeq \nu'$, consistent with a single transition. Outside that range, one should in principle rather rely on a KT fit of the correlation length in the disordered phase to determine the width of the intermediate phase. However, these fits are affected by finite-size effects because, by the time $\xi$ gets very large, it is underestimated on finite systems, hence $1/\xi$ is overestimated, pushing the KT transition deduced from the fit beyond the actual KT transition. 
Still, far enough from the Potts point, and on both sides, there is a clear evidence of a finite intermediate region already for the sizes accessible to our simulations (blue points in Fig. 3(b)). The actual width is expected to lie between the two estimates. 


So, altogether, our results are consistent with a single transition around the Potts point in the range $U>-4.5$ to $V<6$, and with an intermediate phase outside this range, with a very small width below the transition, and an even smaller width above it. If these results are correct, then we expect the presence of two Lifshitz points, one on each side, separating the chiral transition from the incommensurate critical phase\cite{HuseFisher}. Can we be more precise about the location of these points? Unfortunately not. Although we have studied very large systems, finite-size effects are still present, and they prevent us from distinguishing between a very narrow intermediate phase and a direct transition. In fact, even if the results of Fig. 2(g-i) are consistent with a direct chiral transition, we cannot exclude a very narrow incommensurate phase with a width below say $10^{-4}$. So all we can say is that the Lifshitz points (if any) are located between $U=-12$ and the Potts point on one side, and between the Potts point and $V=15$ on the other side. To go beyond these statements one could try to reach even larger sizes, or to study in more detail on the basis of available sizes the crossovers taking place in the vicinity of these putative Lifshitz points. This is left for future investigation.

To conclude, let us briefly compare our findings with previous literature on quantum and classical models. The quantum model of Eq. 1 has been studied in Refs [\onlinecite{fendley}] and [\onlinecite{samajdar}]. Our results confirm some conclusions of both papers, but the overall picture is different from that of both papers. The prediction of Ref [\onlinecite{fendley}] based on Bethe ansatz in the $U\rightarrow - \infty$ that there is a narrow intermediate phase is confirmed by our results for $U$ negative enough, and the suggestion that there could be one above the Potts is now backed by solid numerical evidence. However, the persistence of an intermediate phase up to the Potts point is not supported by our numerical results. The conclusion of Ref. [\onlinecite{samajdar}] that the transition is chiral on the right hand side of the Potts point is supported by our results not too far from the Potts points, but the suggestion that this is true up to $V=+\infty$ is not since we found clear evidence of an intermediate critical phase. We think that the discrepancy comes from the very large system sizes required to detect this incommensurate phase. The simulations of Ref. [\onlinecite{samajdar}] with periodic boundary conditions have the advantage of leading to an accurate estimate of the dynamical exponent through a data collapse assuming that the transition is chiral, but they are limited to sizes that prevent the detection of the incommensurate phase given its very small width. 

Finally, Monte Carlo investigations of the classical chiral Potts model on the square lattice have found evidence of an intermediate floating phase and of a possible Lifshitz point\cite{Selke1982}, but system sizes were too small to study the scaling of the wave vector between the Potts and the Lifshitz point. By contrast, simulations of an extension of Baxter hard-hexagon model away from the integrable point have revealed an exponent $\bar \beta \simeq 0.97$ and a product $\xi q$ that tends to a constant ($q$ being the distance to the ordering wave-vector), pointing to a qualitative difference with Potts universality class\cite{bartelt}, but they did not find evidence of a floating phase. It seems that the identification of the three relevant possibilities (Potts, chiral, and critical intermediate phase) in a single model had not been achieved so far with numerical simulations of classical models. 

{\it Acknowledgments.} We thank Paul Fendley for useful discussions.
This work has been supported by the Swiss National Science Foundation.
The calculations have been performed using the facilities of the Scientific IT and Application Support Center of EPFL.

\bibliographystyle{apsrev4-1}
\bibliography{bibliography,comments}

\begin{thebibliography}{39}
\bibitem [{com()}]{comment1}%
  \BibitemOpen
  \href@noop {} {}\bibinfo {note} {According to the prediction by Fendley et
  al\cite{fendley} in the limit $U\rightarrow \infty$, the critical
  incommensurate phase takes place between $\sigma_{PT}=-13/9$ and
  $\sigma_{KT}=-1.422$, where $U=-3V\left[1+\sigma/V^2\right]$. Using this
  formula for $U=-15$ leads to the estimates $V_{PT}\approx 5.2739$ and
  $V_{KT}\approx 5.2698$, hence to a width $\left. \delta V
  \right|_{U=-15}\approx 0.004$. Our DMRG result $\left. \delta V
  \right|_{U=-15}\approx 0.003$ is consistent with this estimate. In
  particular, it is of the same (very small) order of magnitude.}\BibitemShut
  {Stop}%
\end{thebibliography}

\begin{appendix}

\section{DMRG algorithm in the constrained Hilbert space}

The size of the Hilbert space of hard-bosons grows with the number of sites $N$ as $\varphi^N$, where $\varphi=(\sqrt{5}+1)/2\simeq 1.618$\cite{sierra}, while that of hard-core bosons grows as $2^N$. There is thus a huge gain to be expected numerically if one can write an algorithm that takes the constraint $n_jn_{j+1}=0$ rigorously into account instead of adding a large nearest-neighbor repulsion to penalize configurations with nearest-neighbor bosons. To achieve this, we have proceeded in two steps: (i) We have mapped the hard-boson model onto a quantum-dimer model on a ladder; (ii) We have written a matrix-product state (MPS) version of a DMRG algorithm in which the constraints are explicitly imposed at each step of the algorithm, when constructing the left and right environments, and when constructing the multi-site matrix-product operator. Details will be given in a subsequent paper (Ref. [\onlinecite{chepiga}]). Interestingly enough, the same construction would 
be difficult to implement directly with hard bosons. Indeed, in the quantum dimer formulation, all on-site tensors have a block diagonal structure, which would not be the case for hard bosons.

With this algorithm, and keeping all states with Schmidt values larger than $10^{-12}$, we could reach convergence in a reasonable time on systems with up to 4800 sites keeping up 2200 states and performing up to 12 sweeps. It is worth to mention that, due to low entanglement, the algorithm is unstable when the number of the Lanczos iterations is too low. In particular, in the vicinity of the Pokrovsky-Talapov phase transition the algorithm might be stuck at an excited state that corresponds to several domains of the period-three phase with domain walls between them. The energy acquired by a single domain wall is much smaller than the total energy of the ground-state $\Delta E/E_\mathrm{GS}<10^{-6}$. This makes the convergence to the true ground state in the vicinity of the Pokrovsky-Talapov transition quite challenging for the large systems. However, there are two ways to stabilize the algorithm: either to increase the number of Lanczos iterations to ensure that at each DMRG sweep the lowest-energy state of the effective Hamiltonian is converged to the ground-state, or to diagonalize the effective Hamiltonian simultaneously for a few low-energy states which will push the lowest energy state down to the ground state. While the latter method is rather fast, it is memory consuming since one has to store several Lanczos vectors instead of a single one, so it is suitable for small systems with $N\leq 600$. For larger systems we use the first method and perform up to 150 Lanczos steps or until the energy is converged to $10^{-13}$.

\section{Extraction of the correlation length and of the wave-vector}

\begin{figure}[h!]
\centering
\includegraphics[width=0.45\textwidth]{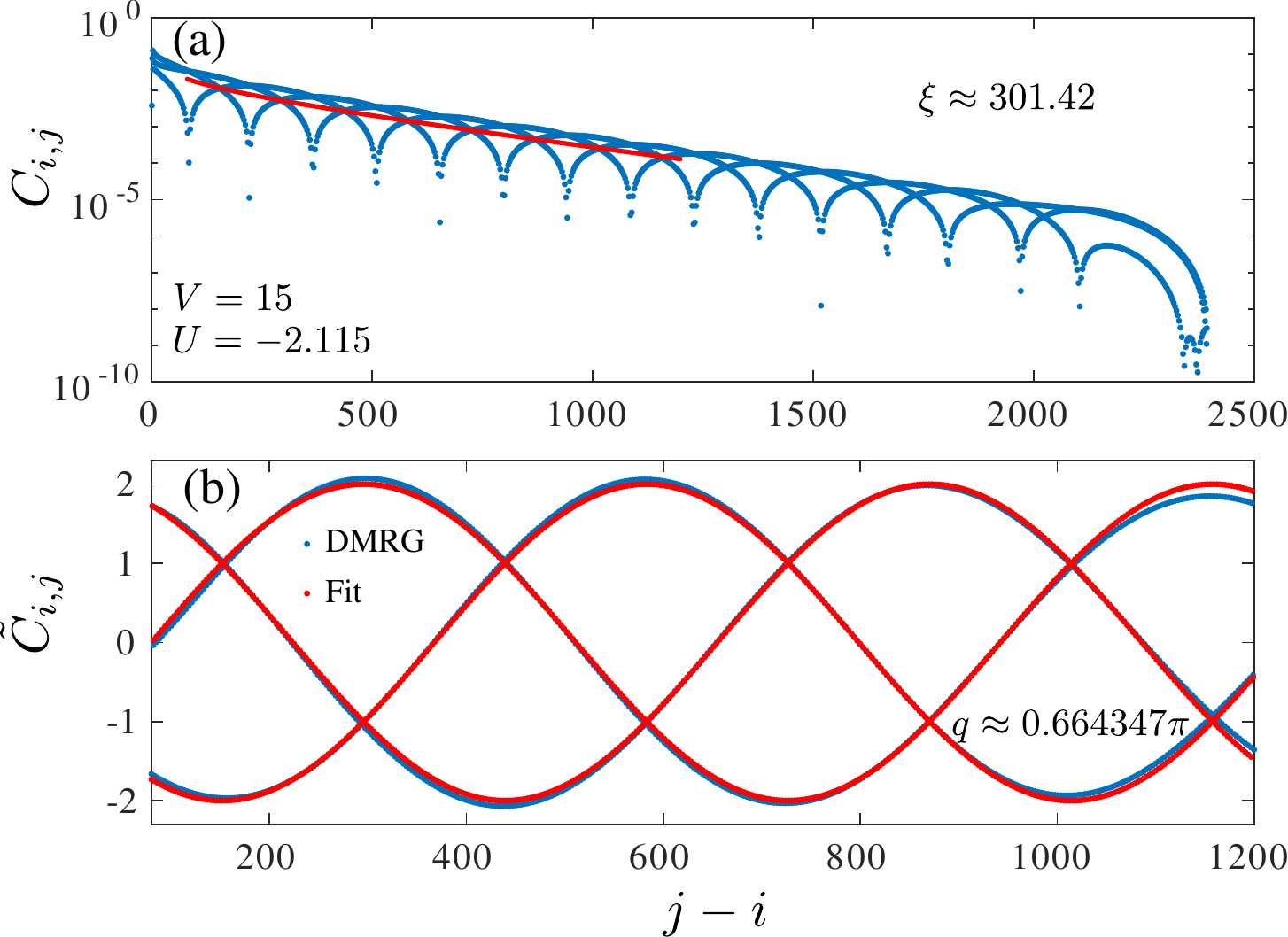}
\caption{  Example of  fit of the correlation function to the Ornstein-Zernicke form. In the first step (a), we extract the correlation length discarding the oscillations. In the second step (b), we fit the reduced correlation function to extract the wave-vector $q$. }
\label{fig:fit}
\end{figure}

In order to extract the correlation length and the wave-vector $q$, we fit the boson-boson correlation function to the Ornstein-Zernicke form:

\begin{equation}
C_\mathrm{OZ}{i,j}\propto \frac{e^{-|i-j|/\xi}}{\sqrt{|i-j|}}\cos(q|i-j|+\varphi_0),
\end{equation}
where the correlation length $\xi$, the wave vector $q$ and the initial phase $\varphi_0$ are fitting parameters. In order to extract the correlation length and the wave-vector with a sufficiently high precision we fit the correlation function in two steps.
 First, we discard the oscillations and fit the main slope of the decay as shown in Fig.\ref{fig:fit}. This allows us to perform a fit  in a semi-log scale $\log C(x=|i-j|)\approx c-x/\xi-\log(x)/2$, that in general provides more accurate estimates of the correlation length on a long scale. Second we define a reduced correlation function
 
 \begin{equation}
\tilde{C}_{i,j}=C_{i,j} \frac{\sqrt{|i-j|}}{e^{-|i-j|/\xi+c}}
\end{equation}
and fit it with a cosine $\tilde{C}_{i,j}\approx a\cos(q|i-j|+\varphi_0) $ as shown in Fig.\ref{fig:fit}(b). The agreement is almost perfect: The DMRG data (blue dots) are almost completely behind the fit (red dots). Since in the first step we did not fit the peaks, but rather average the oscillations out, the pre-factor  $a\neq 1$.

\section{Numerical data along a few selected cuts}

\begin{figure}[h!]
\centering
\includegraphics[width=0.45\textwidth]{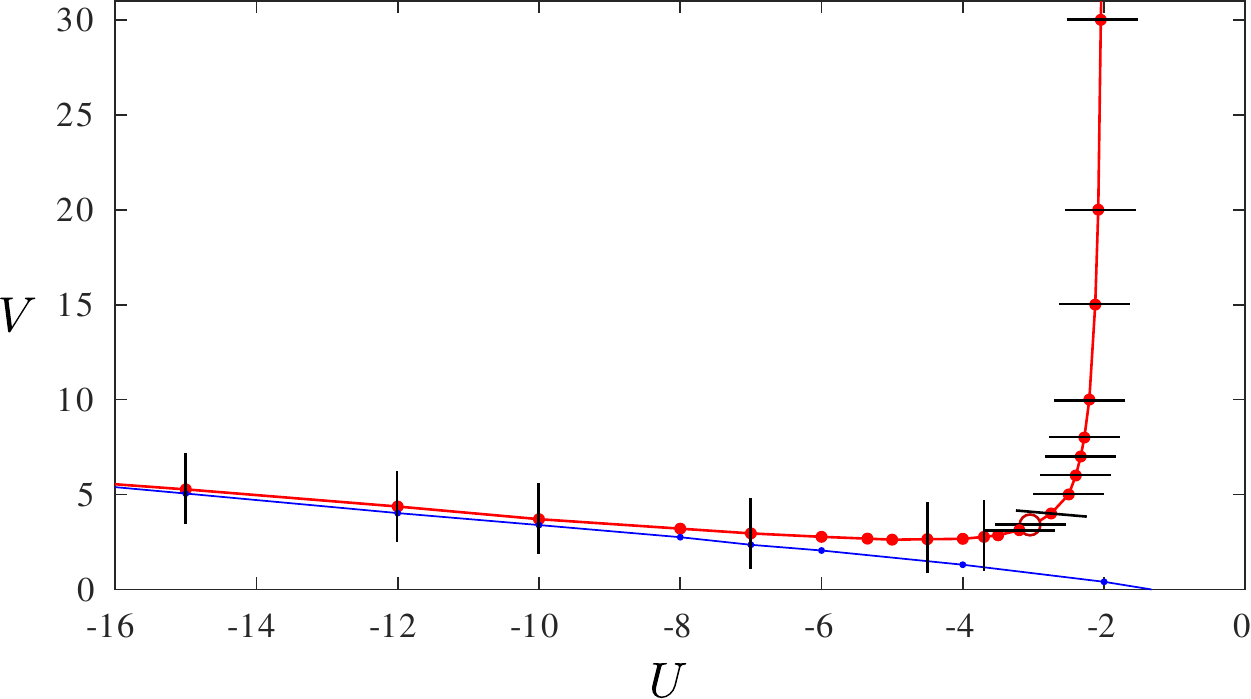}
\caption{ Position of the cuts across the transition between the disordered and period three phases along which we have investigated the properties of the transition in detail. For clarity the length of each cut is enlarged, the actual parameter window used to characterize the type of the transition is smaller than the size of the red circles. The open red circle indicates the critical Potts point.}
\label{fig:cuts}
\end{figure}

In order to probe the nature of the transition along the  boundary between the disordered and period-three phases we extract the correlation length $\xi$  and the wave-vector $q$ along seventeen cuts across transition. The position of the cuts on the phase diagram is shown in Fig.\ref{fig:cuts}.

\begin{figure}[h!]
\centering
\includegraphics[width=0.45\textwidth]{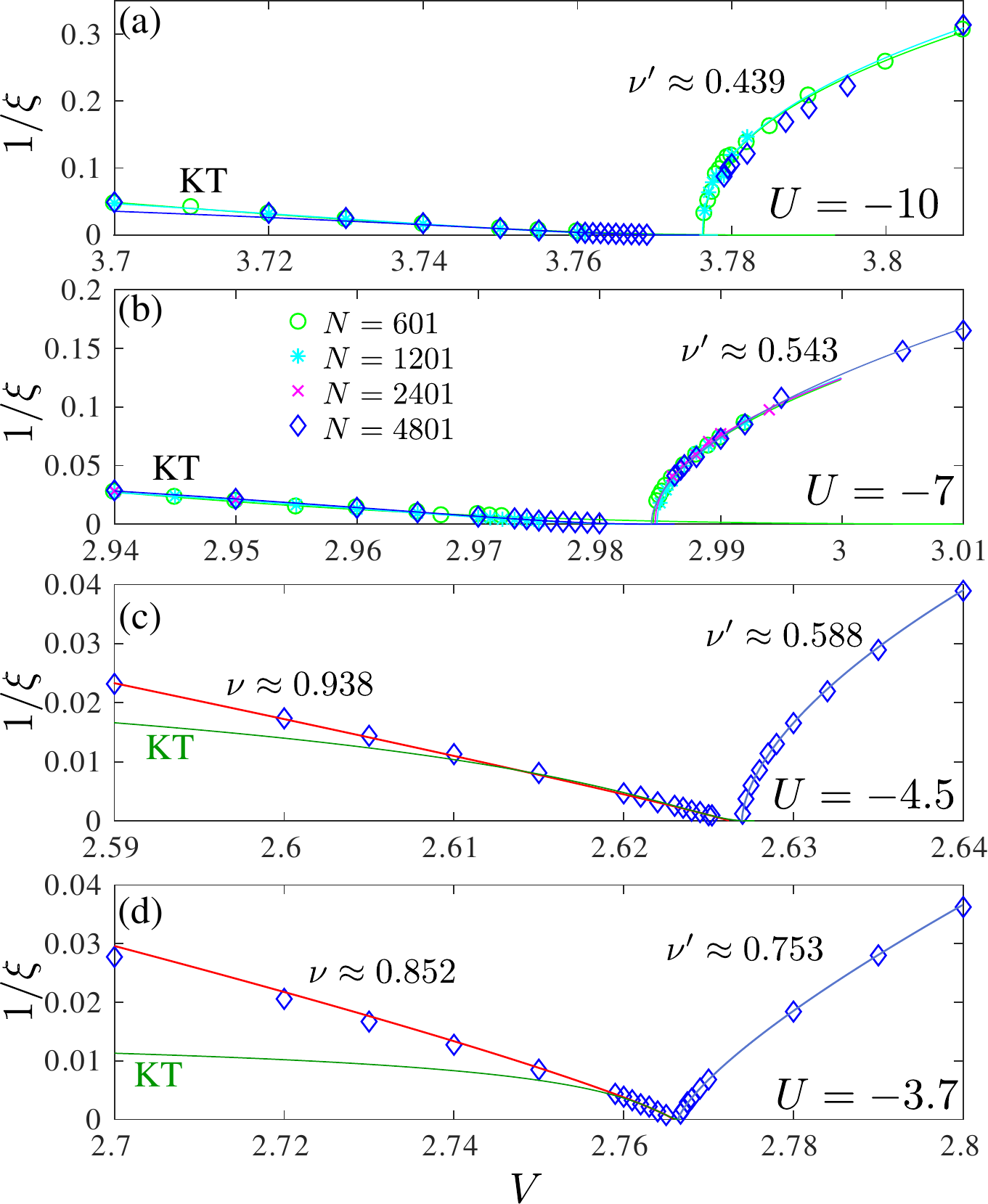}
\caption{ Inverse of the correlation length for a few selected values of $U$ below the Potts point and various values of $V$ around the transition. }
\label{fig:cor_length_below}
\end{figure}

\begin{figure}[h!]
\centering
\includegraphics[width=0.45\textwidth]{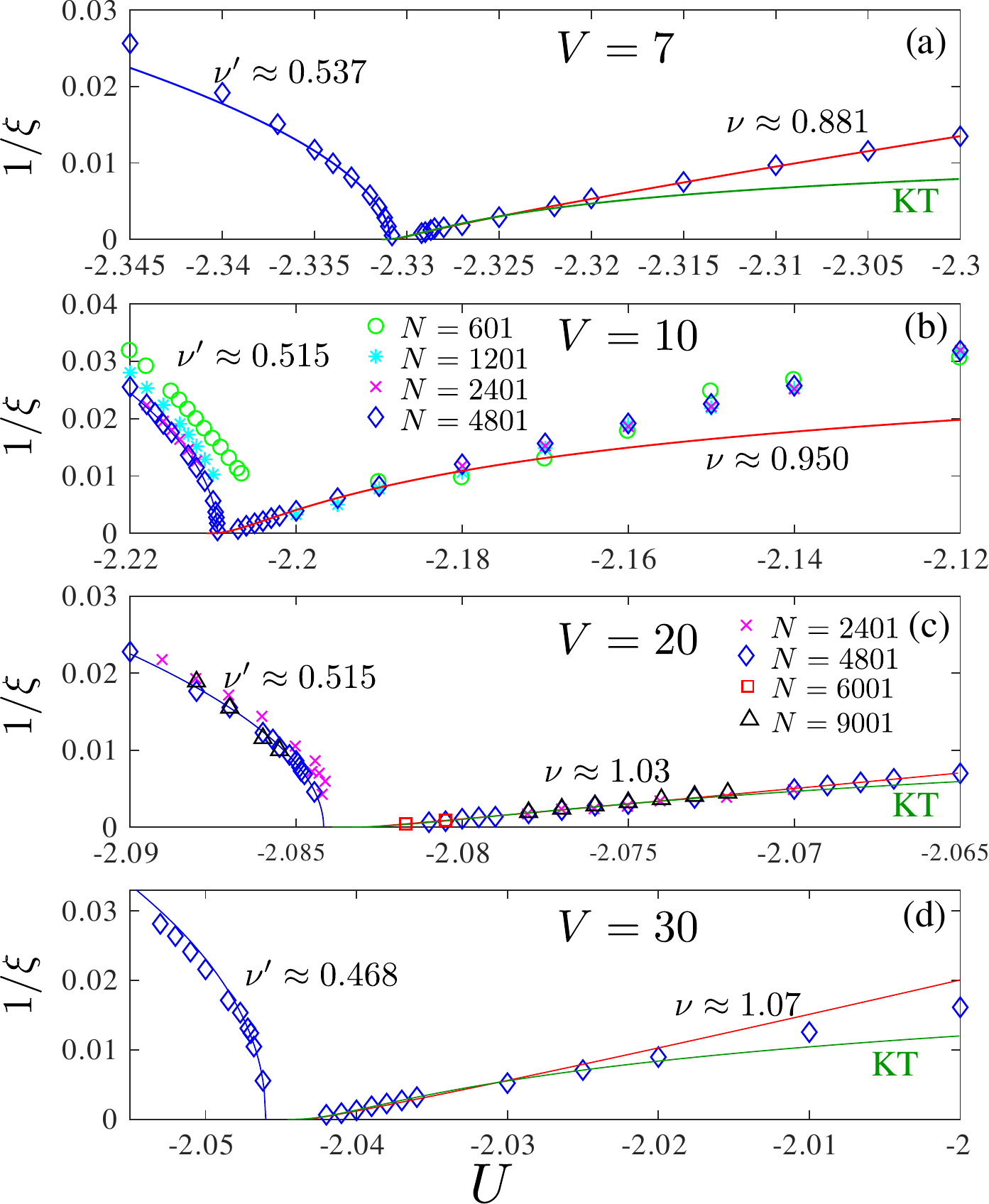}
\caption{ Inverse of the correlation length for a few selected values of $V$ above the Potts point and various values of $U$ around the transition. }
\label{fig:cor_length_above}
\end{figure}

In the main text we provided examples of the scaling of the inverse of the correlation length across the 3-state Potts point, the chiral transition and the intermediate phase. Here we provide more examples of the scaling of the correlation length below (Fig.\ref{fig:cor_length_below}) and above (Fig.\ref{fig:cor_length_above}) the Potts critical point. On the side corresponding to the disordered phase, the curvature changes gradually while moving away from the Potts point. 

\begin{figure}[h!]
\centering
\includegraphics[width=0.45\textwidth]{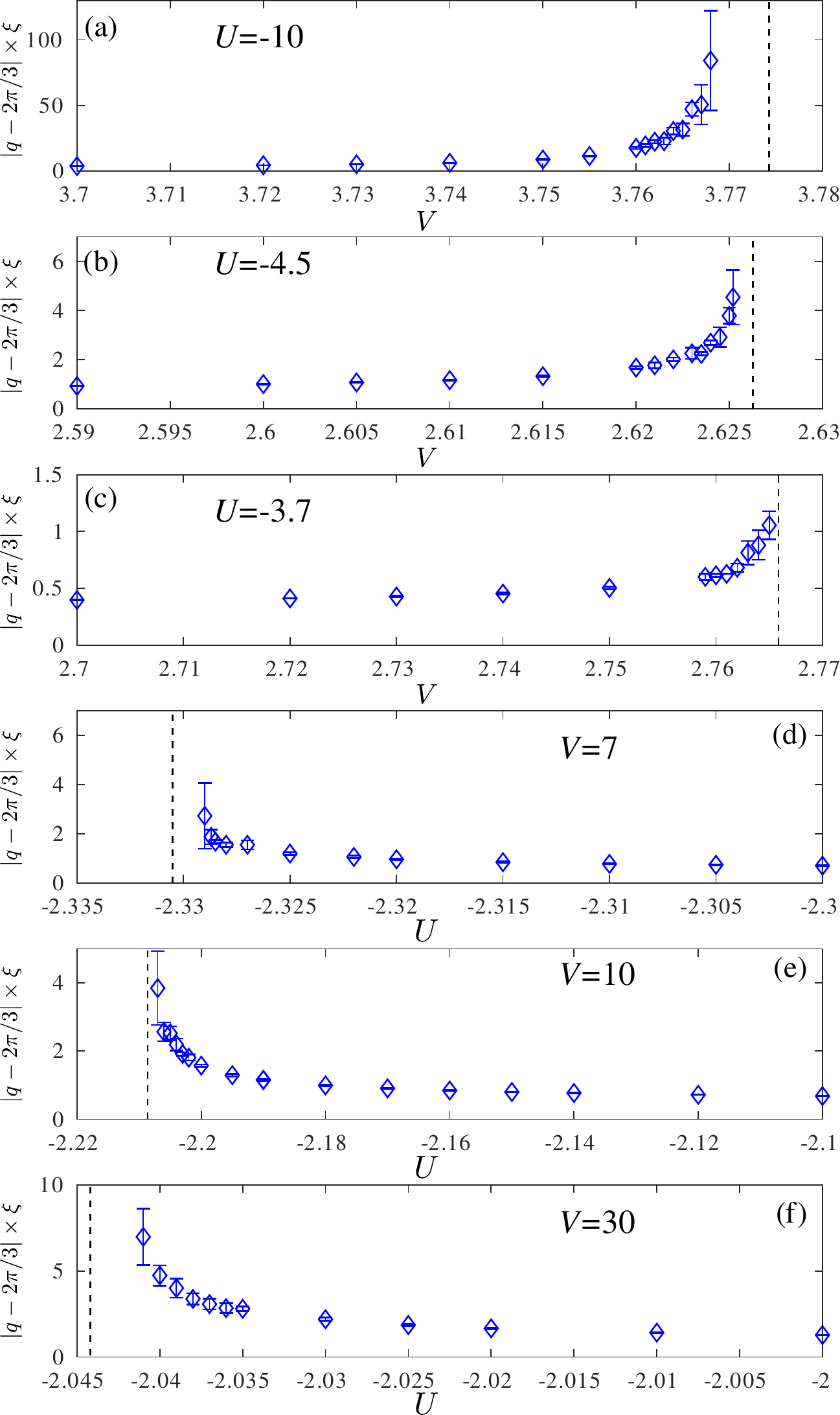}
\caption{ Product $|q-2\pi/3|\times\xi$ across a few selected cuts.}
\label{fig:xi_q_supp_mat}
\end{figure}

Besides, we extract the product  $|q-2\pi/3|\times\xi$ that, as explained in the main text, vanishes at the Potts point, goes to a finite non-universal value at the chiral transition or diverges at the Kosterlitz-Thouless transition. In spite of the noise in the data due to strong finite-size effects close to the critical point, the plots are quite instructive. For instance, a critical phase with divergent $|q-2\pi/3|\times\xi$ can be observed in Fig.\ref{fig:xi_q_supp_mat} panels (a) and (f), while a  finite $|q-2\pi/3|\times\xi$ typical of a chiral transition can be seen in panels (c) and (d). Panels (b) and (e) do not provide conclusive results, presumably because the model lies in the vicinity of the Lifshitz points. 

\section{Central charge of the incommensurate floating phase}

\begin{figure}[h!]
\centering
\includegraphics[width=0.4\textwidth]{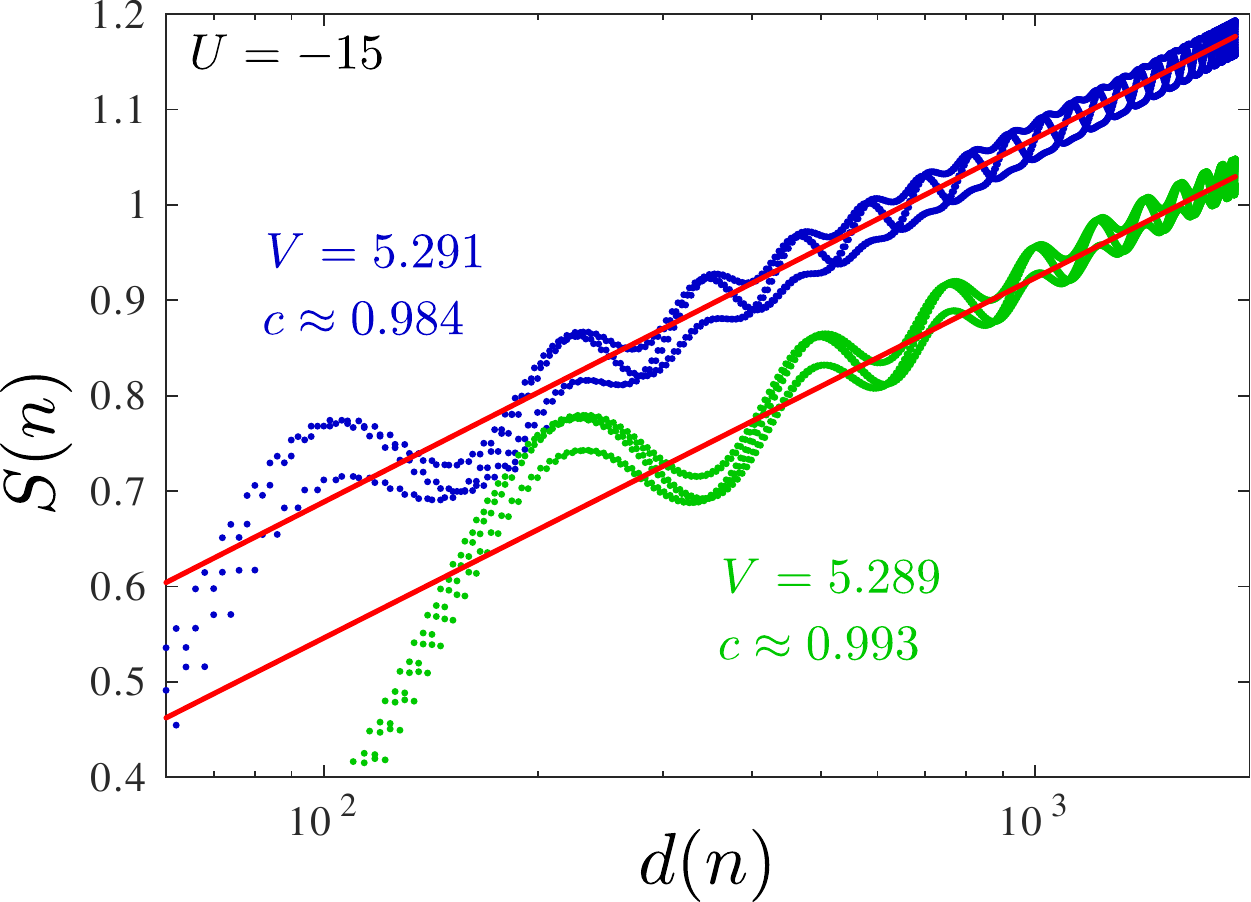}
\caption{  Entanglement entropy as a function of the conformal distance for $U=-15$ and for two values of $V$ within the critical incommensurate phase.  }
\label{fig:u15_cc}
\end{figure}

In order to further confirm that the incommensurate critical phase is in the Luttinger-liquid universality class we have extracted the central charge. According to conformal field theory, the entanglement entropy scales with the conformal distance $d(n)$ as\cite{CalabreseCardy,capponi}:
\begin{equation}
{S}_N(n)=\frac{c}{6}\ln d(n)+s_1+\log g+\zeta\langle {\bf S}_n{\bf S}_{n+1}\rangle,
\label{eq:calabrese_cardy_obc}
\end{equation}
where the last term comes from the Friedel oscillations. In this expression, the conformal distance is given by:
\begin{equation}
  d(n)=\frac{2N}{\pi}\sin\left(\frac{\pi n}{N}\right)
\end{equation}
 For $1\ll n\ll N$, ${S}_N(n)$ oscillates fast with $d(n)$ and with a low amplitude (see Fig.\ref{fig:u15_cc}). So the Friedel oscillation term can be omitted when fitting the entanglement entropy. The resulting central charge agrees within $3\%$ with its value $c=1$ for a Luttinger liquid between the Pokrovsky-Talapov and Kosterlitz-Thouless critical lines.
 
\begin{figure}[h!]
\centering
\includegraphics[width=0.35\textwidth]{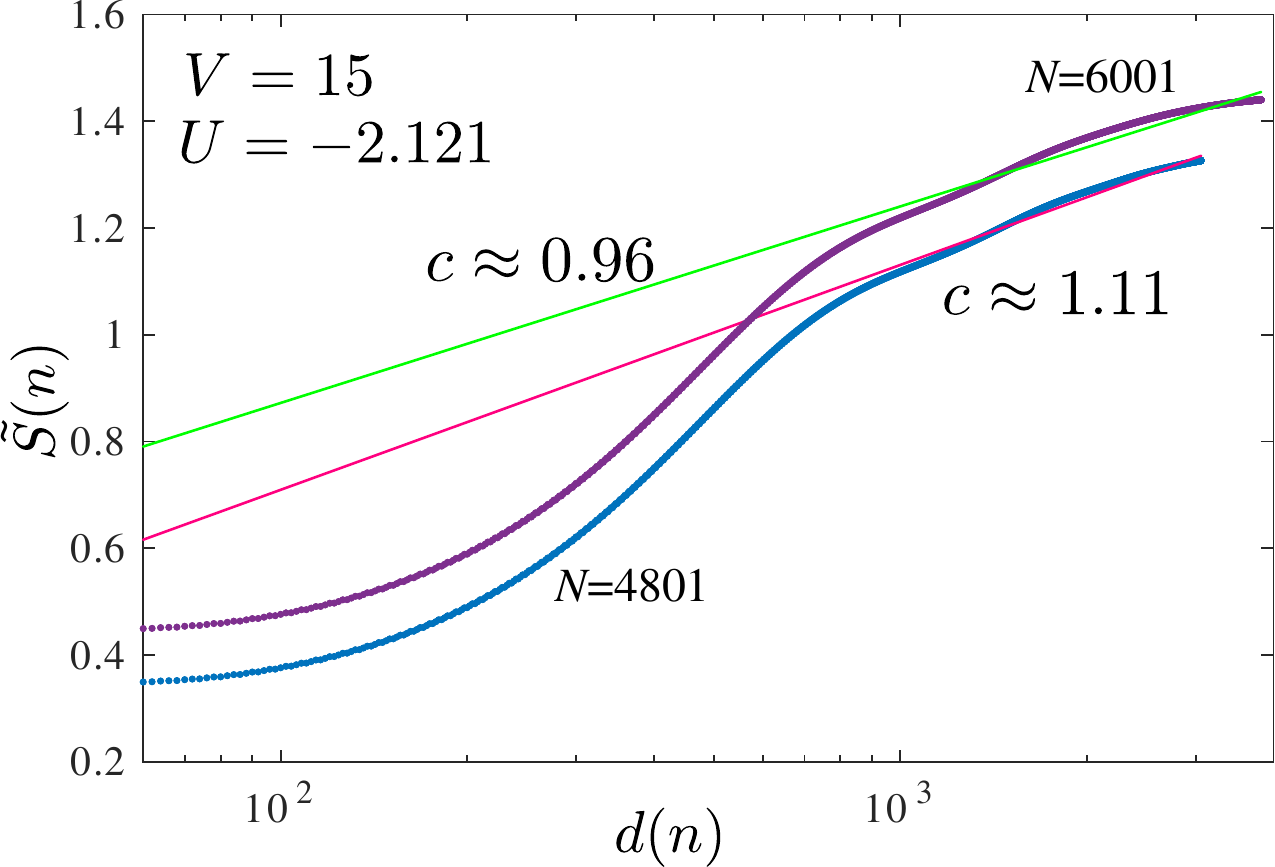}
\caption{  Entanglement entropy with subtracted Friedel oscillations as a function of the conformal distance for $V=15$ and for $U=-2.121$ in the vicinity of the incommensurate phase that starts around $U^c\approx-2.124$. Boundary effects are present up to a distance $d(n)\simeq 400$, beyond which the slope of the entanglement entropy is consistent with a central charge $c=1$.  }
\label{fig:v15_cc}
\end{figure}

Above the Potts point, we were not able to reach  convergence inside the critical phase, but we extracted the central charge at the last available point close to the Kosterlitz-Thouless transition. Since the $q$-vector is close to its commensurate value in this part of the phase diagram, the Friedel oscillations are not fast and cannot be ignored but have to be removed. The scaling of the reduced entanglement entropy $\tilde{S}_N(n)={S}_N(n)-\zeta\langle {\bf S}_n{\bf S}_{n+1}\rangle$ is shown in Fig.\ref{fig:v15_cc}. Close to the edges the behavior is non-universal, however, starting from $n\approx 400$, the scaling is almost linear, and the extracted values of the central charge are consistent with $c=1$.

\begin{figure}[h!]
\centering
\includegraphics[width=0.49\textwidth]{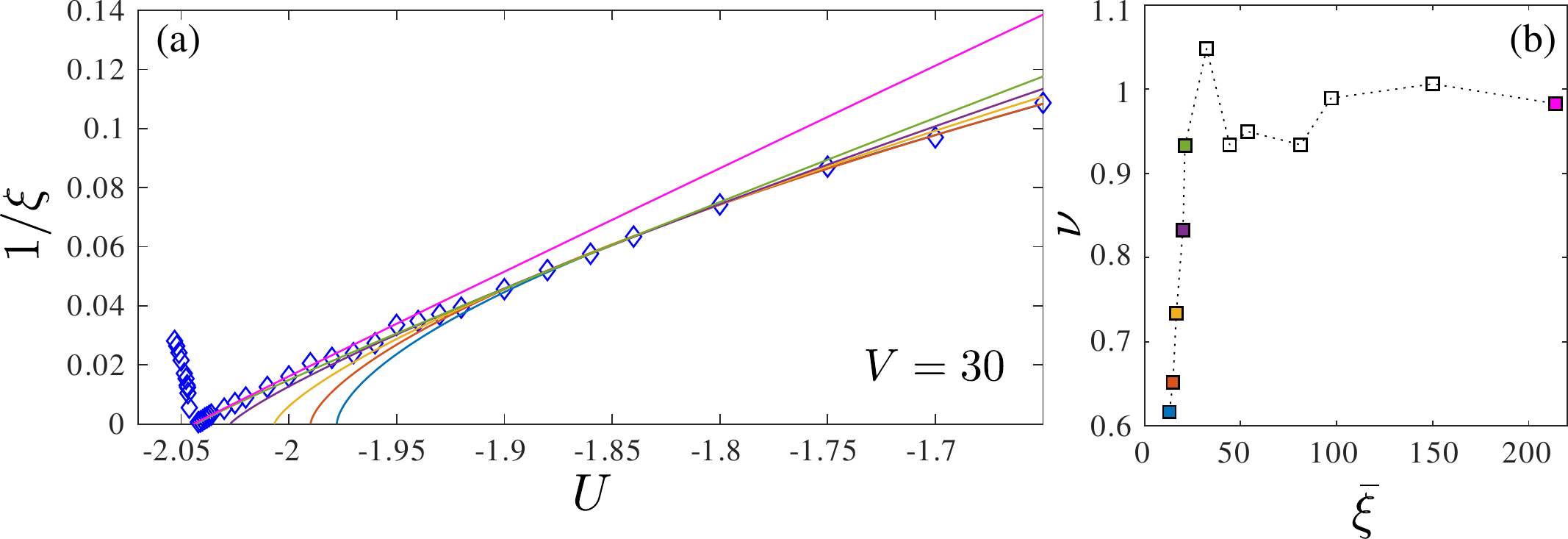}
\caption{(a) Inverse of the correlation length for $V=30$ as a function of U. We fit various sets of 5-8 consecutive points and extract the critical exponent $\nu$. (b) Dependence of the critical exponent $\nu$ extracted in (a) as a function of the mean of the correlation length for the selected points. The colors of the squares correspond to the colors of the lines in panel (a)}
\label{fig:V30_far_from_transition}
\end{figure}

\section{Finite-size suppression of the floating phase}

There are several factors that make the finite-size effects in the vicinity of the floating phase crucial for hard-boson model: (i) the extremely narrow width of the floating phase; (ii) the exponential divergence of the correlation length at the Kosterlitz-Thouless transition; (iii) the fact that the wave-vector $q$ changes very little from its commensurate value inside the critical phase. 

In the present case, open edges favor a particular state and thus correspond effectively to fixed boundary conditions. Therefore the wave-vector changes approximately with steps $2\pi/N$. But the wave-vector $q$ changes very little inside the critical phase. For example, at $U=-15$ at the Kosterlitz-Thouless phase transition $q\approx0.6794\pi$. So, in order to see at least a single helix at this value of  $q$ one has to reach a system size $N\approx2\pi/(q-2\pi/3)\approx157$. 
The finite-size effects are even stronger above Potts point. For example at $V=30$ and at the point at which we identified a Kosterlitz-Thouless phase transition $U^c\approx-2.0441$, the wave-vector $q\approx0.66565\pi$. This implies that the single helix emerges only for $N>2000$.

Besides, the correlation length at the Kosterlitz-Thouless transition diverges exponentially. Assuming that one cannot rely on the data where the correlation length is comparable or even larger than the system size, calculations performed on small systems do not allow to approach the critical point. We illustrate this in Fig.\ref{fig:V30_far_from_transition}, in which we show how the critical exponent changes upon including points with larger correlation length and approaching the phase transition. 
The data points with correlation length $\xi<50$ were obtained on chains of length $N=481$, the points with larger correlation length on chains of length $N=4801$.
As shown in Fig.\ref{fig:V30_far_from_transition}(b), when only small correlation lengths (equivalently small system sizes) are taken into account, the critical exponent is systematically underestimated. This is presumably related to the small exponent obtained  for $V\rightarrow\infty$ in Ref. [\onlinecite{samajdar}].

\end{appendix}

\end{document}